\documentclass[preprint,12pt]{elsarticle}

\usepackage{graphics}

\usepackage{amssymb}

\usepackage{textcomp}
\usepackage{acronym}
\usepackage{color}
\def\Msun{\ensuremath{\mathrm{M_{\odot}}}}

\journal{Comptes Rendus de l'Accad\'emie des Sciences}

\begin{document}

\begin{frontmatter}



\title{Searching for gravitational waves with the LIGO and Virgo interferometers}


\author[Orsay]{M. A. Bizouard}
\author[Hannover,Milwaukee]{M. A. Papa}

\address[Orsay]{Laboratoire de l'Acc\'el\'erateur Lin\'eaire, 
Universit\'e Paris-Sud, CNRS/IN2P3, F-91898 Orsay, France}
\address[Hannover]{Albert-Einstein-Institut, Max-Planck-Institut 
f\"ur Gravitationsphysik, D-14476 Golm, Germany}
\address[Milwaukee]{University of Wisconsin-Milwaukee, WI 53201, USA }

\begin{abstract}
The first generation of ground-based interferometric gravitational wave detectors, LIGO, 
GEO and Virgo, have operated and taken data at their design sensitivities over the 
last few years. The data has been examined for the presence of gravitational wave signals. 
Presented here is a comprehensive review of the most significant results. The network of 
detectors is currently being upgraded and extended, providing a large likelihood for 
observations. These future prospects will also be discussed.
 

\end{abstract}

\begin{keyword}
Gravitational waves \sep neutron stars \sep black holes \sep
supernova core collapse \sep LIGO \sep Virgo.

\end{keyword}

\end{frontmatter}


\section{Introduction}

\acrodef{BBH}{binary black holes}
\acrodef{BNS}{binary neutron stars}
\acrodef{NSBH}{neutron star--black hole binaries}
\acrodef{SNR}{signal-to-noise ratio}

\label{overview}
 
GW observations promise to bring new insight in many phenomena, probing compact 
objects such as neutron stars and black holes, cosmological models of the early Universe and testing general relativity in the strong field regime  \cite{misner:1973,lrr-2006-3,Cornish:2011ys,Li:2011cg}. With the construction and the successful operation 
of the first generation of long arm laser interferometers in the United States 
(LIGO), Italy (Virgo) and Germany (GEO) \cite{Abbott:2007kv,acernese:2008b,Grote:2010zz} the last two decades have been incredibly important for gravitational wave (GW) searches in the $50-1000$ Hz range. 
The upgraded LIGO, GEO and Virgo  
detectors \cite{Harry:2009zz,Accadia:2011}, expected to come online after 2015, along with the Large-scale Cryogenic Gravitational wave Telescope (KAGRA) project in Japan \cite{Kuroda:2011zz} will probe a volume of space 1000 times larger than the first generation of detectors yielding a high probability for multiple observations. For instance, 
40 binary neutron star inspirals per year are expected to be observed with the advanced 
ground-based detectors \cite{ratesdoc} at design sensitivity. However, the first GW detection might be in the $10^{-9}-10^{-8}$ Hz range and come from the analysis of the timing residuals for sets of radio pulsars \cite{PTA2010}.
In the next decade we also hope to see a GW space mission fly, accessing the incredible number of sources in the 
$10^{-4}$ Hz --  0.1 Hz bandwidth.

GW observations will complement the electromagnetic (EM) ones: While EM 
radiation is scattered, absorbed and delayed, due to their weak interaction, GWs carry information from the entirety (especially including the onset) of cataclysmic astrophysical events. Moreover, because of their emission mechanism, GWs carry information about the bulk motion of matter rather than on the differential distribution of charges of the source. Associating Gamma Ray Bursts (GRB) to the gravitational signature of binary neutron star 
and black hole mergers or supernova core collapses will solve an enigma of more than 40 years 
\cite{Zhang:2011rp,Meszaros:2006rc,Nakar:2007}. After many GW detections statistical population studies of the masses of compact binary inspiral systems will address many questions pertaining what is now just an hypothetical population of stellar mass (up to $\sim$ 10 $M_{\odot}$) and intermediate mass (up to $\sim$ 1000 $M_{\odot}$) 
black holes. We also expect to observe a number of GW sources that are EM-invisible: for instance for rapidly spinning neutron stars there exist various channels through which they could to be emitting GWs while appearing EM silent to us.  
The detection of such radiation could bring a wealth of information about the underlying emission mechanism involving crustal deformations, magnetic fields, accretion, and the equation of state of the neutron star interior \cite{prix:2009}.



Individually, each ground-based GW detector is unable to pin down the location in the sky of the source of a short GW signal. A set  
of widely separated interferometers can determine the source position using timing, amplitude and polarization information of the signal seen coherently in the network 
\cite{GuerselTinto1989,Schutz:2011tw}. For this reason since 2007 the LSC (LIGO and GEO) and Virgo have been pooling their data and carrying out their searches jointly. In 2007 the LSC was about to conclude its $\mathrm{5^{th}}$ science (S5) run while Virgo had just started its first science run (VSR1). Since then, another one year long joint science run took place in 2009-2010 
(S6, VSR2 and VSR3). In late 2010 the LIGO detectors shut down for advanced LIGO installation. 
Virgo continued commissioning through late 2011 and improved  
its low frequency sensitivity ($<$ 60 Hz) by a factor of 3. A joint run with the GEO detector was  
conducted in summer of 2011 (S6 for GEO and VSR4 run for Virgo). 
In the following sections, we present the most important results obtained 
with the LIGO, GEO and Virgo data taken during these last years in pursuit of the main sources
of GWs. We will conclude with a brief overview of the astrophysical potential of the network of second generation detectors.

\section{Transient sources and multi-messenger analysis}
Short-lived, or transient, GW signals are thought to be generated in a variety of circumstances. In some cases, for instance the gravitational collapse of a massive star, or the coalescence of two 
neutron stars, GW emission is associated with EM radiation or neutrino
radiation. In other cases the scenarios are more speculative and refer to objects 
that have never been observed as, for example, in the case of stellar mass black hole in binary system or cosmic strings. But all these scenarios could generate transient GWs lasting from a 
few milliseconds up to a few minutes or days in the most extreme cases.

\subsection{Inspiral phase of compact binary systems}

One of the most promising GW sources for interferometric gravitational detectors is 
the coalescence of two compact objects: binary neutron stars (BNSs) , binary black holes (BBHs), or 
a binary neutron star-black hole (NS-BH) system. 
As observed with the binary system PSR 1913+16 discovered in 1974 
\cite{weisberg:2010}, GWs are emitted as the two objects slowly spiral towards each other. They  
will eventually merge and form a black hole.

While binary systems spiral for billions of years, only the signal from their last instants is 
detectable: a few minutes to a few seconds before the merger, when the GW frequency is sufficiently high 
to fall in the detectors' observational bandwidth. The inspiral phase is well described by analytical
post-Newtonian theory until the system reaches the innermost stable circular orbit. Beyond this orbit  
strong gravitational forces dominate and the two bodies merge in a fraction of a second, forming
a single black hole that reaches equilibrium through quasi-normal mode GW emission (ring-down phase).
A BNS inspiral signal will have many cycles in the frequency bandwidth of the LIGO 
and Virgo detectors. For systems with total mass $>$ 100 \Msun~the inner most stable circular orbit is lower than the lowest frequency of LIGO and Virgo detectors and only the merger and ring-down phases 
are visible by these detectors for systems of total mass up to 450 \Msun.
Note, however, that the existence of intermediate mass black holes (systems with total mass between 100 and 450 \Msun) 
is still under debate. Globular clusters are suspected to be the best environment for direct 
collapse (stalled supernova) of very massive first generation stars (population III stars) 
~\cite{Madau:2001sc} or for runaway merging of young stars~\cite{PortegiesZwart:2002jg} leading to 
intermediate mass black holes formation.


\par\noindent
The GW inspiral signal depends on the physical parameters of the binary system: masses, spins and eccentricity. For isolated systems it is generally admitted that the binary systems' orbit has had time to become circular before entering  the detector's sensitivity band, so eccentricity is neglected. Spin is also neglected for low mass binaries, which we expect not to affect detection efficiency. Under these assumptions rather accurate waveforms  can be constructed for signals from the inspiral phase of compact binaries \cite{lrr-2006-9} and used in matched filtering based searches. Only recently the merger phase of some binary configurations has been numerically simulated (see \cite{Centrella:2010zf,lrr-2011-6,Baiotti:2008ra} and references therein). A number of waveforms including inspiral, merger and ring-down phases, without spin have been constructed and used for searching for binary systems with  total mass up to 100 and 450 \Msun \cite{Aasi:2012dr,Virgo:2012ab}. However, these waveforms are subject to large uncertainties for mass ratios $>$ 1:4. Neglecting spin also introduces losses in detection efficiency for intermediate mass black holes searches, and despite the tremendous efforts of the numerical relativity community, template-banks of fully spinning waveforms with spin precession for any mass ratio are not yet available. 
All in all, matched filtering searches are limited to systems of total mass smaller than 100 $\Msun$ and the higher mass parameter space is covered by un-modelled generic transient searches focusing on the merger and ring-down phases ($\sim$ 1 s long signals).

The identification of the source parameters (sky location, masses, spins, polarization, 
orientation, ...) will be the step immediately following a GW detection. Markov Chain Monte Carlo or Nested Sampling techniques have been exploited to do this \cite{Christensen:1998gf, Veitch:2009hd} exploiting waveforms with fully precessing spins and the most recent theoretical investigations \cite{Raymond:2009cv}.

\subsection{Sources of bursts of GWs}
As seen in the previous section, as the total mass of the compact binary system increases, the inspiral part of the GW signal happens at lower and lower frequencies leaving only the merger and ring-down phase in the sensitive band of the instruments. The merger phase is rapid (seconds or less) and despite the advances in numerical relativity it is not well modeled. For this reason intermediate mass binary black hole mergers are also searched using  methods that do not rely on a specific waveform model, but rather on excess-power techniques whose detection efficiencies do not depend on specific matching of wave-shapes. This type of methods were originally developed in the broader context of searching for generic short-lived signals. The spectrum of transient searches includes searches associated with EM or neutrino counterparts as well as broad all-sky, all-times searches: Unforeseen transient phenomena involving, say, newly formed black holes or neutron stars accreting matter in highly magnetized 
media could well be associated with the emission of bursts of gravitational radiation.
Known mechanisms for the emission of GW bursts include the gravitational core collapse of a massive star, neutron star glitches and cosmic strings. In the remainder of this section we will briefly review these mechanisms.

The different mechanisms involved in stellar collapse have been studied for more than 
40 years, and GW emission is predicted in different phases: 
it is anticipated during the collapse itself through the aspherical bulk mass 
motion (core bounce phase). In the post-bounce phase, convection instabilities inside and 
above the proto-neutron star can develop, driven by entropy gradients, neutrino heating and 
standing accretion shock instability (SASI) mechanisms (see \cite{Ott:2008wt,lrr-2011-1} 
and references therein). These phenomena may produce GWs lasting up to few seconds. 
After a supernova explosion or a collapsar (black hole formation) a certain fraction of 
the ejected material can fall back, heating the proto-neutron star and exciting oscillation modes that 
generate GWs (g-mode pulsations and r-mode instabilities). 
Furthermore, when a black hole is formed, accretion-induced ringing may generate GWs. 
Current state-of-the-art numerical simulations take into account many non axisymmetric effects 
in 3D, attempt to model micro-physics (neutrino transport, electron capture, realistic 
equation of state) and include rotational dynamics and magnetic fields. Still, we are quite far from having a complete catalogue of all possible waveforms. For instance, GWs emitted during the core bounce exhibit a sharp peak in the waveform for most of the initial conditions and energy up to a few $\mathrm{10^{-8}}$ $\Msun\mathrm{c^{2}}$ can be emitted \cite{Dimmelmeier:2002bm}, but the exact signal shape depends on the rotational profile 
and the nuclear equation of state. Consequently, for the searches, we consider only the generic broad features of the waveforms in the literature. 

Some pulsars occasionally show sudden discontinuities (glitches) in their EM activity followed by gradual recoveries 
that may last days or years. Soft Gamma-ray Repeater (SGR) 
flares \cite{Mereghetti:2008je}, anomalous X-ray pulsars (AXPs) as well as radio glitches are examples of this and are thought to be consequences of star-quakes  
or of angular momentum transfer between a solid crust and a more rapidly rotating loose 
component of the neutron star interior \cite{lrr-2008-10}. During such glitches normal 
modes (mainly f-modes) could be excited and generate GWs.  


Cosmic strings predicted in symmetry-breaking phase transitions~\cite{vilenkin:1994} 
or cosmic super-strings in string theory inflation models form a dynamical network. When 
cosmic strings interact, they generate loops whose oscillations can lead to cusps (sharp bends)
which move at the speed of light and generate burst of GWs~\cite{Damour:2001bk}. The GW amplitude 
depends on the tension of the string. The (super-)string density depends on the the loop size, 
the string tension and an additional parameter, the reconnection probability, which in the 
case of super-strings is smaller than 1. 


\subsection{Multi-messenger analyses}
Knowing the location of a putative source and the time of occurrence of a GW event thanks to the existence of an external, EM or neutrino, trigger restricts the GW search parameter space and improves the sensitivity of triggered searches by roughly a factor 2 with respect to all-sky, all-time searches. GRBs, long and short, magnetar flares (SGRs and AXPs), pulsar glitches and high energy neutrinos events have been used as triggers for short-lived-GW searches. 
In the remainder of this section we will briefly describe the physical process underling each external trigger and describe how it could be associated with GW production. We will conclude by considering multi-messenger analyses in which the GW search triggers follow-up observations rather than being triggered by an EM or neutrino event.

GRBs are the brightest EM phenomena in the Universe and are likely associated 
to cataclysmic stellar events with a typical energy release of $10^{51}$ ergs.
Long GRBs are related to the core collapse of massive stars ~\cite{1999ApJ...526..152F,Metzger:2011aa} which, we have seen in the previous section, can generate GWs. 
Short GRBs are thought to be generated when a BNS or a NS-BH 
merge forming a black hole and a dense accretion disk \cite{nakar07,Rezzolla:2011da}. As we have seen in the previous section BNS and NS-BH mergers are among the prime targets for LIGO/Virgo. A small fraction of short 
GRBs may come from SGR giant flares \cite{Nakar:2005hs}. 
GWs emitted by galactic SGRs could be detected by the first generation of interferometric 
detectors. 

Single high energy neutrino (HEN) events, although not yet directly observed, could well be emitted in processes that also produce GWs. Since the HEN and GW detectors have uncorrelated backgrounds, requiring a coincidence in time and source location between the HEN and GW events could significantly reduce the false alarm rate of a joint  observation and hence improve the sensitivity of the search \cite{Baret:2011nu}. Plausible scenarios for joint HEN-GW emission include again GRBs, as well as micro-quasars and magnetars \cite{Ando2012,Bartos:2011aa}.

In core collapse supernova, EM waves are emitted in optically thin regions far from the 
core, which provide limited information about the early instants of the core collapse. 
Low energy neutrinos and GWs are emitted deep inside the core and would provide insight in supernova 
mechanism. Moreover, the coincident detection of GWs and low energy neutrinos in the absence of any EM counterpart would provide evidence for the existence of so called failed supernovae, which could occur at a rate comparable with that of visible supernovae \cite{failedSN}. 



Unlike EM telescopes that can only observe some part of the sky at any given time, GW detectors see the whole sky all the time\footnote{this is strictly true only with a network of GW detectors spread over the world, where blind sky regions of one detector can be covered by other detectors}. This means that GW observations could be used to alert EM telescopes of interesting locations to point: The putative source location of GW triggers identified in close to real time could be passed on to telescopes for follow-up observations. Being able to measure the full light curve of a supernova or GRB prompt emission and afterglow 
in all wavelengths would help understand all the mechanisms involved in the 
cataclysmic event, including an investigation of the host galaxy at the time of the 
event. Since directional estimates are still challenging for a 3 GW detector network (error 
box of up to tens of square degrees for signals just above detection threshold 
\cite{Cavalier:2006rz, Fairhurst:2009tc}), large field of view telescopes will be required for this 
during the first GW observations. In 2009-2010,
a first campaign of $\sim$ 3 months with an heterogeneous set of partners (optical telescopes, 
X-ray and radio instruments) brought much valuable insight for the preparation of
advanced detectors era GW alert system \cite{Abbott:2011ys, Abadie:2011bn}, pointing out especially
the loose accuracy of the GW source sky localization.



\subsection{Transient search results}

To date, no GW transient signal has been discovered in LIGO, GEO or Virgo data. As a
consequence, upper limits on the rate of GW events or on the amount of energy released 
have been derived which can be compared to astrophysical predictions.

\begin{figure}[h]
  \begin{center}
    \includegraphics[angle=0,width=0.75\linewidth]{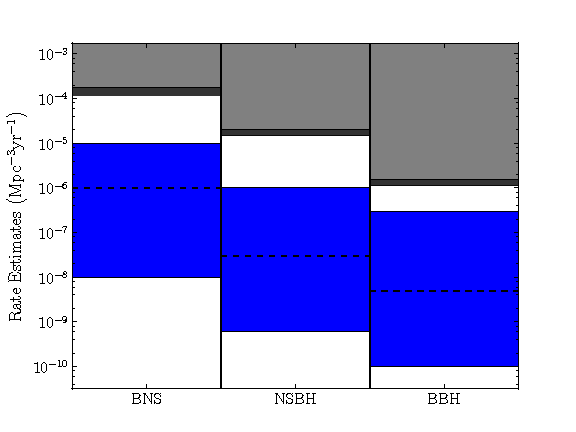}
    \caption{Comparison of upper limit rates for BNS, NS-BH and BBH systems \cite{Abadie:2011nz}. The dark gray regions show the most recent upper limits obtained with LIGO and Virgo data. The new limits are up to a factor of 1.4 improvement over the previous results (light grey regions). The lower (blue) regions show the spread in the astrophysically predicted rates, with the dashed-black lines showing the ``realistic'' estimates \cite{ratesdoc}.}
    \label{fig:CBC}
  \end{center}
\end{figure}

Predictions of the rate of coalescences of compact binary systems are rather uncertain, spanning 1 or 2 orders 
of magnitude in each direction from the most likely prediction. They are based on 
observations of binary neutron stars in our galaxy or on simulations (population synthesis) 
of systems involving black holes, which have not been observed yet. The rates 
are then extrapolated to our local Universe considering the stellar birth rate in nearby 
spiral galaxies. Following \cite{ratesdoc} we will use the following figures as realistic predictions 
for the number of expected coalescences per Mpc$^3$ and per year for various binary systems (also shown in Figure \ref{fig:CBC}): 
$\mathrm{10^{-6}~ Mpc^{-3}~ yr^{-1}}$ for BNS systems, 
$\mathrm{3\times10^{-8}~ Mpc^{-3}~ yr^{-1}}$ for a 10 $\Msun$ NS-BH system and 
$\mathrm{5\times10^{-9}~ Mpc^{-3}~ yr^{-1}}$ for a 10 -- 10 $\Msun$ BBH system. 

The most recent data (from the S6,VSR2-3 runs) have been searched for signals from compact binary coalescences of systems with total mass up to 25 $\Msun$. The waveforms used did not include spin and were terminated just before the two objects merge. The typical distance up to which the search could have detected these systems was 40 Mpc, 80 Mpc and 90 Mpc for the BNS, NS-BH and BBH systems respectively. 
Because no detection was made upper limits on the rates of such astrophysical events were placed at 
the 90\% confidence level: 
$\mathrm{1.3 \times 10^{-4}~ Mpc^{-3}~ yr^{-1}}$ for BNS systems,
$\mathrm{3.1 \times 10^{-5}~ Mpc^{-3}~ yr^{-1}}$ for a NS-BH system and 
$\mathrm{6.4 \times 10^{-6}~ Mpc^{-3}~ yr^{-1}}$ for a BBH system as shown 
in Figure \ref{fig:CBC} \cite{Abadie:2011nz}. These numbers fold in as priors the results 
from previous runs. The binary neutron star rate upper limit is still two orders of magnitude higher than 
what is likely expected. The detection efficiency of the search is lower for spinning systems resulting in upper limits on the rate of NS-BHs and BBHs that are $\sim 16\%$ larger than for non spinning systems. Spin is not expected to significantly change the waveform of BNS systems visible by the instruments. 

For binary systems with
a total mass higher than 25 $\Msun$ (BBH system regime) the merger phase contributes a non negligible fraction 
of the total signal-to-noise ratio. For this reason in the search \cite{Aasi:2012dr} for systems with total mass between  25 $\Msun$ and 100 $\Msun$ the template-bank utilizes waveforms that cover both the inspiral, merger and ring-down phases of the signal. The resulting maximum distance reach on S6,VSR2/VSR3 data is 300 Mpc for a 20 -- 20 $\Msun$ system, which is  significantly higher than for any system in the low-mass search reported above \cite{Abadie:2011nz}. In the search post-Newtonian inspiral waveforms are connected to the merger and ring-down waveforms with an effective 
one-body approach tuned on numerical relativity waveforms \cite{Buonanno:2007pf,Pan:2011gk}. Spin effects are not included in the template waveforms. Including spin in the search in the most general way is well outside what is currently possible, due mainly to the lack of numerical relativity simulations over the broad parameter space of possible configurations. There exist, however, waveform models for spin configurations aligned or anti-aligned with the orbital angular momentum, at least for mass ratios between 1 and 4. Figure 7 of \cite{Aasi:2012dr} shows that the average reach of the search for such systems varies by about 10\% with respect to the non-spinning population. Upper limits on the rate of coalescences of such systems were derived as a function of the component masses. For example, for systems of 50$\Msun$+50$\Msun$ the rate upper limit is $\mathrm{0.7 \times 10^{-7}~ Mpc^{-3}~ yr^{-1}}$ and the distance reach well exceeds 200 Mpc. We refer the reader to Figure 5 of \cite{Aasi:2012dr} for the complete set of results over the different mass bins. Expectations for the rate of coalescences of high mass systems are rather uncertain. However there exist predictions \footnote{based on two observed tight binaries with individual masses of $\sim$ 15 $\Msun$ and $\sim$ 20 $\Msun$.} for the rate of  $\mathrm{3.6 ^{+5.0}_{-2.6} \times 10^{-7}~ Mpc^{-3}~ yr^{-1}}$ for stellar mass binary systems \cite{Bulik:2008ab,Belczynski:2011qp}. In the corresponding mass bin the GW search constraints the rate to $\mathrm{5.9 \times 10^{-7}~ Mpc^{-3}~ yr^{-1}}$ which already includes part of the predicted rate-space, but spin effects were not considered in the search.

For binary systems
of higher mass the only part of the coalescence waveform in the instruments' band is the merger and ring-down. A  simple ring-down search carried out in 2004 LIGO data yielded an event rate
upper limit of $\mathrm{3.2 \times 10^{-5}~ Mpc^{-3}~ yr^{-1}}$ for binary total mass between 
100 and 400 $\Msun$. In S5,VSR1 data a search for intermediate mass black hole systems in the total mass range 100 and 450 $\Msun$ and with mass ratio up to 1:4 was performed. This search is an excess power search, it does not rely on a specific waveform model and coherently combines the data from the different detectors \cite{Virgo:2012ab}. The largest effective
range, 241 Mpc, is achieved for a $88+88~ \Msun$ system and in this mass bin the rate upper limit is $\mathrm{1.3 \times 10^{-7}~ Mpc^{-3}~yr^{-1}}$. The upper limit on the event rate, averaged
over all considered masses, is $\mathrm{9 \times 10^{-7}~ Mpc^{-3}~yr^{-1}}$ which is just an order of 
magnitude higher than the most constraining rate upper limit. As globular clusters (GC) are the most likely hosts
of intermediate mass black holes, the event rate can be converted into a limit on the astrophysical source
density: $\mathrm{3\times 10^{-6}~ GC^{-1}~ yr^{-1}}$. This is still 3 order of magnitudes higher 
than the astrophysical predictions \cite{ratesdoc}, which is anyway rather uncertain as intermediate
mass black holes formation is still under debate.

The most sensitive all-sky generic transient signal search to date was carried out on the S6,VSR2-3 data and would have been able to detect with $50\%$ efficiency a burst ($<$1s long) of $2.2 \times 10^{-8}~\Msun c^2$ of gravitational radiation with significant energy content around 150 Hz at the distance of 10 kpc \cite{Abadie:2012rq}. The total observational time for this search amounts to 1.74 yr. Upper limits on the rate of different types (waveforms) of events in the frequency range between 64 Hz to 5 kHz,  as a function of signal amplitude at the Earth are derived. For signals strong enough (root-sum-square strain amplitude $\mathrm{h_{rss}} \sim $ several $10^{-20}$ $\mathrm{strain \over {\sqrt{Hz}}}$) the rate upper limits all tend to $\mathrm{1.4~ yr^{-1}}$.
The results can also be interpreted as limits on the rate density of GW bursts assuming 
a standard candle source isotropically distributed\footnote{this is a rough estimate when the 
distances are within our local universe.}.
For instance for sources emitting $\mathrm{10^{-2}~ \Msun c^2}$ around 150 Hz, the rate is bound to 
$6 \times 10^{-3}~ \mathrm{Mpc^{-3}~yr^{-1}}$. 

\begin{figure}[h]
  \begin{center}
    \begin{tabular}{cc}
    \includegraphics[angle=0,width=0.45\linewidth]{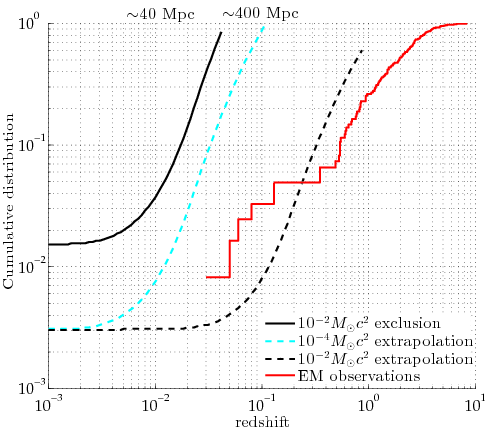} &
    \includegraphics[angle=0,width=0.45\linewidth]{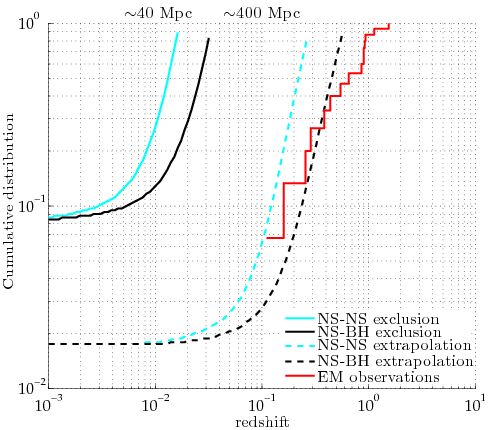} \\
    \end{tabular}
    \caption{Cumulative distribution 90\% confidence level exclusion as function of redshift from the analysis of 151 long GRBs (left) and 26 short GRBs (right) for different hypothesis on the GRB progenitors (smooth curves). The dashed curves are extrapolations assuming a factor 10 in sensitivity is brought by advanced LIGO-Virgo detectors. The red stairs show the cumulative distribution of measured redhsifts for GRBs \cite{Briggs:2012ce}.}
    \label{fig:GRBs}
  \end{center}
\end{figure}

As explained above, short GRBs are probably associated with the coalescence of NS-NS or NS-BH systems. On the other hand long GRBs have been associated with massive star core collapse. Even though only for $\sim$ 10\% of the observed GRBs there exist redshift measurements it is believed that the typical distance for GRB progenitors be on the Gpc scale, well beyond the reach of the first generation of GW detectors. However based on observations of a few GRBs at much closer distances (e.g. GRB 980425 at 36 Mpc) and on indications on the existence of a class of sub-luminous long GRBs, one cannot exclude the chance that there might be a detectable gravitational signal in association with a GRB trigger. For this reason searches for GWs from binary inspirals and searches for generic GW transients have been carried out specifically around the time of occurrence of 350 GRBs observed by gamma-ray satellite detectors since 2003 \cite{Briggs:2012ce, Abadie:2010uf, Abbott:2009kk, Abbott:2008zzb}. 
The Swift and Fermi satellites have been the primary sources  
of the most recent GRB triggers targeted with LIGO-Virgo data\footnote{the list of the analyzed GRBs, 
their origins and their main characteristics are given in publications 
\cite{Briggs:2012ce, Abadie:2010uf, Abbott:2009kk, Abbott:2008zzb}.} taken when at least two GW detectors were operational. No GW detection resulted from any of these searches. The most recent search \cite{Briggs:2012ce} excludes the existence of progenitors within a certain distance from Earth, which is different for every GRB, assuming an emission of $10^{-2}~\mathrm {\Msun c^2} $ at 150 Hz in GWs. The median distance for this set of GRBs is 17 Mpc. From the null results of the binary coalescence signal searches the corresponding median distance values are 16 Mpc and 28 Mpc for NS-NS and NS-BH systems respectively.
Further, if one assumes all GRB progenitors to emit the same amount of GWs, the set of GRBs can also be 
considered as a whole. Constraints on the cumulative number of GRBs as function of the redshift are shown 
in Figure \ref{fig:GRBs} where a constant co-moving density rate of GRBs up to a cosmological distance
beyond which all GRBs are at infinity is assumed. Extrapolations at the foreseen advanced GW detectors sensitivities reach redshift values in the range of current short GRBs observations. For long GRBs, one needs to assume a rather optimistic GW emission ($10^{-2}$ \Msun $\mathrm{c^2}$) in order to reach observed redshifts. 

\par\noindent
GW observations were able to exclude certain galaxy associations for two GRBs, GRB 051103 \cite{Abadie:2012bz} and GRB 070201 \cite{Abbott:2007rh}, at least if these GRBs were due to BNS mergers. The position obtained by the interplanetary 
network of satellites for GRB 051103 had a $3-\sigma$ error region overlapping with the outer disk of the M81 
galaxy (3.63 Mpc). GRB 070201's position was 1.1 degree away from the center of Andromeda-M31 
(770 kpc). A GW binary coalescence signal from either M81 or M31 would have been well within the sensitivity reach of the operating LIGO detectors. The absence of a GW detection following searches for such signals allowed to exclude with high confidence the association between the GRBs and a BNS merger in their apparent host galaxies. 
However it was not possible to exclude SGRs in the apparent host galaxies as the progenitors of these observed short GRBs ($\sim$ 15\% of the short GRBs are accounted for as SGRs \cite{Nakar:2005hs}).

GW signals around the times of observed X-rays emission of 6 galactic SGRs and anomalous X-ray pulsars 
have also been searched. The amount of GW energy to be expected and the mechanism of emission are still poorly 
understood. However, some theoretical models predict a GW energy emission as large as 
$\mathrm{10^{48} - 10^{49}}$ ergs \cite{Corsi:2011zi}. Among all magnetars that have emitted 
x-rays when LIGO, Virgo and GEO were taking data, SGR 1806\textminus20 was the most energetic 
transient observed in our galaxy ($\mathrm{\sim 4 \times 10^{46}}$ ergs). Given the astrophysical 
uncertainties, no specific waveforms were assumed in the search, and 90\% upper limits on the 
amount of GW energy emitted by these objects have been set considering white noise bursts with 
frequencies $\sim$ 100 Hz as representative of GW signals from low frequency oscillation modes and considering damped sinusoids with 
frequencies $>$ 1 kHz as representative of GW signals from f-modes. The most constraining upper limit
has been obtained for SGR 0501\textminus4516 (1 kpc): it ranges from $\mathrm{3 \times 10^{44} ~ergs}$ to $\mathrm{1 \times 10^{47} ~ergs}$ for un-modelled low frequency GW emission and for f-mode 
damped sinusoid waveforms with 1090 Hz central frequency, respectively \cite{Abadie:2010wx,Abbott:2009zd}.
It is interesting to note that values around $\mathrm{10^{45} ~ergs}$ are comparable to the 
EM energy of giant flares\footnote{it is however unknown whether EM
and GW energies are correlated.}. Furthermore the f-mode limit of $\mathrm{1 \times 10^{47} ~ergs}$ is in the range predicted by \cite{Corsi:2011zi}. 

The Vela pulsar is one of the pulsars with the least stable spin phase evolution: Superimposed on the steady decrease in the rotation velocity of this neutron star, astronomers observe a few times a year sudden increases in its spin. This could be a consequence of star-quakes that adjust the crust of the neutron star or rearrangement of the interior material, and are expected to excite normal mode oscillations that may emit GWs.
On August $\mathrm{12^{th}}$ 2006 a glitch amounting to a relative increase of the rotational frequency of $\mathrm{2.62 \times 10^{-6}}$ was measured. This glitch happened when the two collocated LIGO detectors at Hanford were taking data. A search was performed for ring-down GW waveforms from the quadrupolar (spherical harmonic order $\ell=2$) f-mode oscillations with frequencies between 1 and 3 kHz and damping times between 50 and 500 ms. No GW signal
was observed and 90\% confidence level upper limit on the peak strain amplitude of ring-down 
signals have been set between $\mathrm{6.3 \times 10^{-21}}$ and $\mathrm{1.4 \times 10^{-20}}$, depending
on which of the $2\ell +1$ harmonics is excited. This peak strain value corresponds to a GW energy range of $\mathrm{5 \times 10^{44}}$ ergs to
$\mathrm{1.3 \times 10^{45}}$ ergs \cite{Abadie:2010sf}.

High energy neutrinos (HENs) may be emitted in various circumstances in association with GWs and although many scenarios, such as GRBs and SGRs, also involve EM emission that is already used as trigger for targeted GW searches, it is not excluded that there might be joint GW-HEN emission with scarcely or not at all visible EM counterpart (for example associated with chocked GRBs). The first GW search triggered by HEN candidates was carried out using HEN triggers from the Antares experiment on S5/VSR1 GW data \cite{AdrianMartinez:2012tf}. The astrophysical significance of the null result that ensued is limited, however this is the first ever GW-HEN search and it paves the way for searches in the advanced GW detector era when, with GW sensitivity increases of up to a factor of 10 it is expected that, if not detections, an interesting range of rate upper limits will be probed.

A dedicated search for GWs from cosmic super-string cusps has been performed on S4 data. It has not 
revealed any GW signal and upper limits on the event rate have been set and used to exclude portions of the 
cosmic strings parameter space (string tension, loop size and reconnection probability). Big
Bang Nucleosynthesis indirect exclusions are more stringent, but the GW search sensitivity obtained
with more recent data is surpassing the indirect bounds in some portion of the parameter space \cite{Abbott:2009rr}.

\section{Continuous wave searches}
\label{sec:CW}
\subsection{Sources}
Neutron star are amazing objects for GW astronomy. Not only are they the most promising 
and well predictable source of GWs when they are part of a coalescing binary system, but spinning
neutron stars can also radiate themselves a detectable amount of GWs. 

In the frequency band of ground-based GW detectors (10~Hz -- 2~kHz), different emission 
mechanisms can lead to long-lasting, quasi-monochromatic GWs (see section III of \cite{Abbott:2006vg} and references therein). Among these, non axisymmetric deformations are considered the most reliable mechanism, at least for isolated systems. These deformations in the crust can be generated by elastic stress \cite{Bildsten:1998ey, Ushomirsky:2000ax} or by magnetic fields  \cite{cutler:2002nw}. 
Mechanisms other than crustal distortions include GW driven instabilities of normal oscillation modes  (r-mode \cite{Andersson:1997xt} and f-modes Chandrasekhar-Friedman-Schutz \cite{Friedman:1978hf} instabilities), or free precession, when the star's rotational axis does not coincide with its symmetry axis.

In the case of non-axisymmetric deformations, the gravitational wave frequency is twice the spin frequency and the amplitude of the emitted GW is proportional 
to the deformation which is expressed in terms of the ellipticity $\epsilon$\footnote{$\epsilon={{I_{xx}-I{yy}}\over {I_{zz}}}$, where the $I$s are the moments of inertia of the star and the spin axis is assumed to lie in the $z-$direction.} of the distorted object. Predictions for $\epsilon$ are 
highly uncertain. The maximum deformation that a crust can support depends on the 
neutron star equation of state through the shear modulus of the inner crust and on the crust's structure (crystalline or amorphous), through its breaking  strain \cite{lrr-2008-10}. Possible breaking strain values span several orders of magnitude, the largest ones ($\sim 10^{-1}$) corresponding to perfect crystals. Shear modulus values of the inner crust also span a few orders of magnitude, with solid strange-quark star models yielding the highest shear modulus values. Adding it all up maximum values of the ellipticity range from several $10^{-7}$ to a few $10^{-4}$. 

But what are the ellipticity values in actual pulsars ? This question does not have a clear-cut answer and it cannot be dismissed that a certain fraction of the neutron stars have 
negligible axial quadrupole moment. This is supported by observations of extremely low period derivates of millisecond pulsars that indicate that these recycled pulsars have little equatorial asymmetry ($\epsilon < 10^{-8}$). Simulations of isolated 
spinning neutron star populations including gravitational and electromagnetic torques 
\cite{Wade:2012qc} 
provide insight in the type of constraints that even non-detections of continuous GW signals will place on neutron star population parameters in the advanced detector era. 

Accreting neutron stars in binary systems have  a natural way to maintain large crust deformations, through the accretion spot. 
In fact it has been suspected for a long time that GW emission 
could balance the accretion torque and explain why spin frequency of known accreting neutron stars are never 
approaching the breaking limit \cite{Bildsten:1998ey,Andersson:2000pt}. However the actual search for signals from binary systems is significantly complicated by the additional unknown orbital parameters that one would need to search over \cite{Watts:2009gk}.

\subsection{The detection problem}
In the rest frame of the source and in absence of glitches or star quakes, continuous GWs have a constant amplitude and are quasi-monochromatic, with a 
slowly decreasing frequency (spin-down). Due to the relative motion between the source and the Earth, GWs are received at the detectors with a Doppler modulation. Additionally, due to the non uniform sensitivity of the detector across the sky, the signal amplitude also appears to be modulated. If one knows the frequency and frequency derivatives of the incoming GW signal and its origin in the sky, all the modulation effects can be removed and, broadly speaking, the detection problem broils down to that of detecting a sinusoidal signal, hence to integrating the data for long enough to accumulate sufficient signal-to-noise ratio\footnote{The real situation is actually more complicated because of other unknown parameters (i.e. the inclination angle of the system, the polarization of the GW, the phase) but these may be maximized over or marginalized over and for the purpose of this discussion we can neglect them. }. This is conceptually what is done when targeting known objects for which the spin phase evolution, position and orbital parameters (if part of a binary) are known.

For the vast majority of the known systems with GW frequency in the range of ground-based detectors we do not expect to be able to make detections. This statement is based on comparing the typical strain (or amplitude) sensitivity of our search with the spin-down limit for the same system. The spin-down limit is given by the amplitude of a continuous GW that carries away energy at the observed rotational kinetic energy loss rate. 
It is the maximum GW amplitude that we can expect from a system for which we know the frequency evolution. For most known systems in the LIGO and Virgo frequency bands the spin-down limit is well below the values that our searches could detect.

However there are many more neutron stars in our Galaxy than we know of ($\sim 10^9$), and the most interesting searches for continuous GWs aim at discovering a system that is spinning at the right frequency and is bumpy enough and close enough to be detected. The main challenge with this type of search is that one has to explicitly search over all parameters : position, frequency and spin-downs, if the system is isolated, and also over orbital parameters for systems in binaries. The most sensitive techniques that combine the data coherently on longer and longer timescales $T$ cannot be used because the resolution in parameter space increases rapidly with the observation time ($\propto T^5$ for searches including only first order spin-downs), thus making the problem computationally prohibitive. We hence resort to hierarchical methods that iteratively identify at lower resolution regions in parameter space more likely to harbor a signal and then refine the search with higher resolution and sensitivity around such regions. The specific combination of the parameters that characterize the various stages is tuned depending on the data and on the parameter space that one wants to search. In the next section we discuss representative results of various types of searches for continuous wave signals.



\subsection{Results}

In the absence of a detection we place upper limits on the intrinsic GW strain amplitude $h_0$. Depending on the emission mechanism such upper limit can be recast in terms of other physical quantities. If the emission is due to non-axisymmetric deformations then \cite{Jaranowski:1998qm}
\begin{equation}
h_0=\mathrm{{16\pi^2 G\over c^4}{\epsilon I f^2\over r}}
\end{equation}
where $\mathrm{I}$ is the moment of inertia, $\epsilon$ the ellipticity, $\mathrm{f}$ the GW frequency and $r$ the distance of the source. 
If one knows the distance to the putative source, as in the case of searches that target known objects, it is straightforward to see how upper limits on $\mathrm{h_0}$ may be recast as upper limits on the ellipticity at a fixed nominal value of $\mathrm{I}$ (typically chosen $\mathrm{10^{38}~ kg m^2}$ ) or as an exclusion region in the $\mathrm{I-\epsilon}$ plane, allowing $\mathrm{I}$ to vary. 

When the upper limits stem from a search that does not target any specific object or an object whose distance is unknown, the $\mathrm{h_0}$ upper limits can be recast as lower limits on the distance of the source having assumed that it is emitting GWs because of a non-axisymmetric deformation and that it is doing so at the spin-down limit for some value of the spin-down. 

Under the assumption that the GW emission stems from r-modes, Eqs (23) and (24) of \cite{Owen:2010ng} allow to convert $\mathrm{h_0}$ upper limits in upper limits on the dimensionless r-mode amplitude $\alpha$ for fiducial values of the source parameters such as mass, radius and equation of state\footnote{Note however that the conversion from $\mathrm{h_0}$ upper limits derived from detectability studies of a specific search on a specific population of signals is correct if the polarization angles of the injected population were uniformly distributed.}.

\begin{description}
\item {Targeted searches (known position and frequency):}
The ephemerides of many pulsars are regularly updated thanks to optical, X-ray and $\gamma$-ray pulsar 
observations. Such ephemeris are then used to determine the putative signals' Doppler and amplitude modulations employed by all methods 
developed to search for known pulsars in GW data: i) a matched-filter method \cite{Jaranowski:1998qm}\cite{Cutler:2005hc}  ii) a time-domain heterodyne search which generates Bayesian posteriors on the parameters associated with the signal (amplitude, source 
orientation, initial phase and polarization angle) \cite{Dupuis:2005xv} 
iii) a Fourier-domain determination of the signal at the 5 frequencies (carrier plus 2 pairs of sidebands) 
at which the signal power is split by the amplitude and phase modulation due to the sidereal variation of 
the detector's beam pattern functions \cite{Astone:2010zz}. The first continuous wave search performed with LIGO data targeted one pulsar, J1939+2134 
($f_{rot}=642 Hz$) using data of 2002 \cite{Abbott:2003yq}. Since then, the number of 
known pulsars searched for in each new data set  has increased steadily while more pipelines have been developed and refined. This, combined with more sensitive data and longer data sets has led  
to more constraining upper limits over the time \cite{Abbott:2004ig,Abbott:2007ce,Abbott:2009rfa}.
With S5 data the lowest strain amplitude upper limit is $2.3 \times 10^{-26}$ achieved 
for PSR J1603\textminus7202 and the lowest limit on the ellipticity is $7.0 \times 10^{-8}$ for 
PSR J2124\textminus3358. The $h_0$ upper limits values from our GW observations are more constraining than the spin-down limits for two young pulsars: the Crab pulsar ($f_{rot}$=29.78 Hz) and the Vela pulsar ($f_{rot}$=11.2 Hz). For the Crab the 95\% confidence upper limit on the strain amplitude is $2.0 \times 10^{-25}$, yielding a 2\% upper limit value on the fractional rotational 
energy loss through GW emission \cite{Abbott:2008fx, Abbott:2009rfa}. 
For Vela  the strain amplitude limit is $2 \times 10^{-24}$, bounding to less than 35\% the amount of 
rotational energy \cite{Abadie:2011md} lost in GWs. Albeit the non-detection per se is not unexpected since the ellipticity values corresponding to the spin down limits are rather high (a few $10^{-4}$ and $10^{-3}$ for the Crab and Vela pulsars respectively), the constraints on the fraction of energy loss in GWs are informative measurements that could not be obtained by other means.

\item {Directed searches (known position but unknown frequency):}
These searches are somewhere in-between the targeted searches described above and the all-sky searches for completely unknown objects that will be described next. By directed searches we mean searches aimed at a given direction in the sky where a compact object, most of the time a non pulsing neutron star, may be emitting strong continuous GWs. This is for instance the case of the supernova remnant Cassiopeia A ($\sim$ 3.4 kpc) which is suspected to host a compact central object born less than 400 years ago. Under the assumption that its spin frequency evolution has been largely governed by GW emission, an indirect limit on the GW strain amplitude, $\mathrm{h_0^{age}}$, can be derived which only depends on the age of the object
\begin{equation}
\mathrm{h_0^{age}=1.2\times10^{-24}({3.4~ kpc\over D}) \sqrt{({I_{zz}\over {10^{38} kg~m^2}})({300 years \over \tau})}}
\end{equation}
with $\tau$ being the age of the object \cite{Wette:2008hg}. In the absence of any observational information on the rotation period of the central object, a 200 Hz wide frequency band was searched  for continuous signals from this object in S5 data \cite{Abadie:2010hv}. The search band was chosen so that the minimum detectable strain amplitude across that band would be lower than the $\mathrm{h_0^{age}}$ limit. 
 No significant GW candidate was identified and 95\% confidence upper limits on the strain amplitude ($7 \times 10^{-25}$ at 150 Hz), ellipticity ($4\times 10^{-4}$ at 100 Hz), and r-mode amplitude ($5 \times 10^{-3}$ at 300 Hz) were derived \cite{Abadie:2010hv}. 

The low mass X-ray binary Sco X-1 (2.8 kpc), the brightest X-ray source in the Galaxy, is expected to be 
also a strong GW emitter. The detection of continuous GWs from this system would allow to test the torque balance mechanism of accreting neutron stars. The rotation frequency of Sco X-1 is unknown making the search over frequency, spin-down and orbital parameters computationally very challenging. A 20 Hz wide frequency band was searched in LIGO 2003 data using of only few hours of the most sensitive data.
The upper limit on the strain amplitude is few orders of magnitude higher than the torque balance prediction, which is $\sim 3 \times 10^{-26}$ for a source emitting GW at 540 Hz \cite{Abbott:2006vg}. 

\item {All-sky, all-frequency searches (objects not previously known): }

As described above, all-sky, all-frequency searches for unknown neutron stars are the most challenging of the continuous GW searches carried out to date. These searches typically do not begin before all the data to be analyzed has been acquired. The core search lasts of the order of a year and the post processing takes a comparable amount of time. This means that results from this type of analyses lag behind the results for other types of signals. The most recent results come from two searches on S5 data \cite{Abadie:2011wj,Aasi:2012fw}. In \cite{Abadie:2011wj} we add incoherently 30-minutes long power spectra over the entire duration of the run, spanning a frequency range between 50 and 800 Hz. This is a fast-turn-around of results type of search. 
In \cite{Aasi:2012fw} the searched frequency range extends to 1200 Hz (the higher frequencies are the most computationally expensive), one year of S5 data is considered, split up in 121 30-hour segments where the data from the two detectors is coherently combined. This is a very compute-intensive search and it leveraged the compute power of Einstein@Home\footnote{\tt{http://einstein.phys.uwm.edu.}}, a volunteer computing project where members of the general public donate compute cycles from their personal computers in order to help with a scientific endeavor. Both methods (Powerflux and Einstein@Home) set upper limits on the strain amplitude as function 
of the GW frequency which are comparable. For instance, the best 90\% CL upper limit achieved by 
Einstein@Home is $7.6 \times 10^{-25}$ at 152.5 Hz in agreement with the expected sensitivity of the analysis. The reach of the search extends to a few kpc around 150 Hz.

\end{description}

\begin{figure}[bph!]
  \begin{center}
    \includegraphics[angle=0,width=0.9\linewidth]{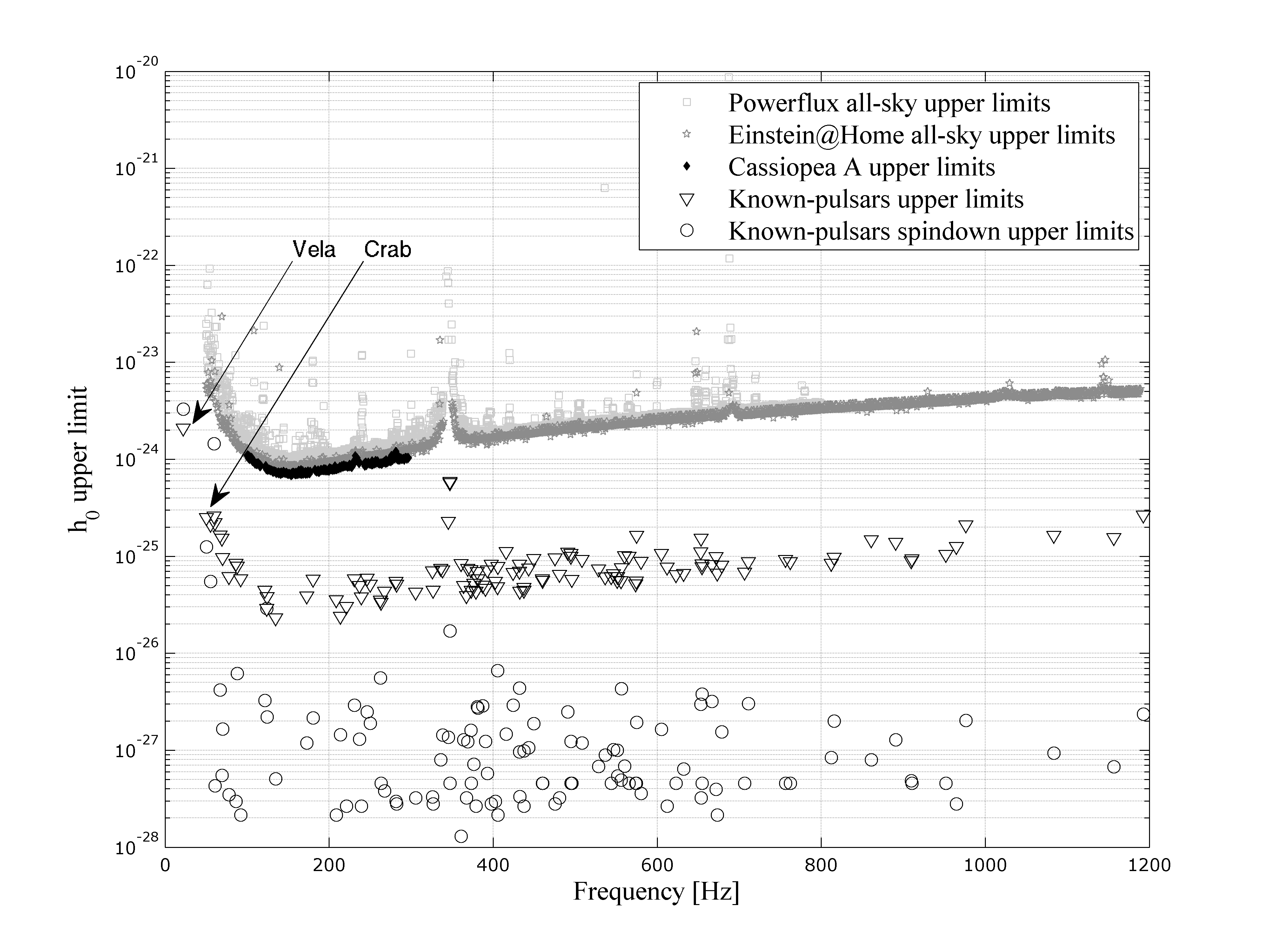}
    \caption{The indirect spin-down upper limits (open circles) for the known pulsars are below the measured values (open triangles) for all pulsars but the Crab (at about 60 Hz) and the Vela pulsar (at about 22 Hz). The $h_0$ upper limits from these searches are the lowest compared to the ones from other searches because the parameter space is much smaller : virtually one template versus $10^{13}$ or $10^{18}$ templates for the Cassiopeia A (black small stars) and the  Einstein@Home (dark grey crosses) searches, respectively. This affects the sensitivity in two ways 1) through the trial factor, i.e. the largest random fluctuations increase over a larger set of independent trials and this is why a signal needs to be larger in order to be detected with the same confidence over a larger parameter space 2) the most sensitive and computationally intensive coherent search techniques can be employed when the same search has to be repeated fewer times. The Powerflux (light grey open squares) search examined more than $10^8$ points in parameter space but with a non coherent method.}
    \label{fig:CW}
  \end{center}
\end{figure}

The most recent results of  continuous wave searches are summarized in Figure 
\ref{fig:CW}.   

\section{Stochastic gravitational wave background searches}
\subsection{Sources}
A cosmological and astrophysical stochastic background of GWs (SGWB) is expected from the incoherent 
superposition of GWs emitted by many sources. Cosmological sources include amplification of quantum 
vacuum fluctuations during inflation~\cite{Grishchuk:1974ny}, 
electro-weak phase transitions~\cite{Apreda:2001us}, 
pre Big Bang models~\cite{Brustein:1995ah,Buonanno:1996xc} and 
cosmic (super-)strings~\cite{Kibble:1976sj,Damour:2004kw,Siemens:2006yp}. 
Most of these processes took place in the early stages of the universe making GWs a unique way to 
probe primordial evolution. An astrophysical SGWB may be generated by the concurrent emission of a large population of sources such like compact binary coalescence and merger, core collapse supernovae or rotating neutron stars.

\begin{figure}[bph!]
  \begin{center}
    \includegraphics[angle=0,width=0.85\linewidth]{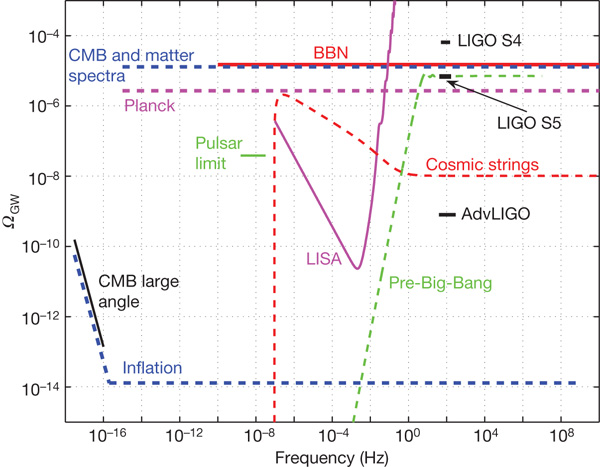}
    \caption{Comparisons of different SGWB search upper limits on the GW spectrum $\Omega_{GW}$ 
and indirect bounds \cite{Abbott:2011rs}.}
    \label{fig:SGWB}
  \end{center}
\end{figure}

A SGWB is described in terms of the GW energy density
\begin{equation}
\Omega_{GW}(f) = \frac{1}{\rho_c}\frac{d\rho_{GW}}{d\mathrm{ln} f}
\end{equation}
where $\rho_c$ is the critical energy density of the universe and $d\rho_{GW}/d\mathrm{ln} f$ 
is the GW energy density per logarithmic frequency interval.
Predictions on $\Omega_{GW}$ as well as indirect phenomenological and experimental bounds  are illustrated
in figure \ref{fig:SGWB} which is taken from \cite{Abbott:2011rs}. The most stringent bound comes from Big Bang nucleosynthesis (BBN) and cosmic microwave background (CMB) measurements, and it depends on number $N_{\nu}$ of neutrino families at the time of the BBN ~\cite{Maggiore:1999vm} : 
\begin{equation}
\Omega_{BBN} = \int \Omega_{GW}(f) d\mathrm{ln} f < 1.1 \times 10^{-5} (N_{\nu}-3)
\end{equation}
Primordial nucleosynthesis provides an upper bound $N_{\nu}-3 < 1.4$.
The CMB data can be used to derive a limit on the maximal SGWB energy density at the time of 
photon decoupling~\cite{Smith:2006nka}

\begin{equation}
\Omega_{CMB} = \int \Omega_{GW}(f) d\mathrm{ln} f < 1.3 \times 10^{-5}
\end{equation}
These are the two phenomenological upper bounds that can be directly compared to direct LIGO-Virgo 
measurements. 

\subsection{Results}
Since the signal is a stochastic process the search is made by cross-correlating detector strain 
amplitudes (the GW amplitude is much smaller than the instrumental noise and the signal is expected to be
continuous). For cosmological SGWB sources one assumes that the SGWB is isotropic and entirely described 
by a spectral power density whose frequency dependence is parameterized by a power law with index $\alpha$,
$\Omega_{GW}(f) = \Omega_\alpha (f/f_r)^{\alpha}$, where $f_r$ is an arbitrary reference frequency. 
On the contrary for astrophysical sources 
it makes sense to search a background with a spatial distribution, as if it was coming from a point in 
the sky. 

The sensitivity of each pair of detectors depends on the noise power spectrum of each detector and on the
observing geometry (detectors location and orientation and spatial distribution of the background). Having
co-located and co-aligned detectors like the two Hanford LIGO detectors maximizes the search sensitivity
over the full frequency band of the detectors, but taking into account correctly the presence of correlated 
noise that could cancel, corrupt or mime the measurement of a SGWB is a serious issue. 
On the contrary the LIGO Hanford and 
Livingston pair presumably do not have correlated noise and benefit from their good relative alignment that
yields good sensitivity in the low frequency region where the detectors' sensitivity is the highest. The isotropic search carried out in the $\mathrm{41.5 - 169.25 ~Hz}$ frequency band with 
S5 data of this pair of detectors set a direct upper limit on $\Omega_{0}$ at 95\% confidence 
level of $6.9 \times 10^{-6}$~\cite{Abbott:2011rs}. A SGWB of this strength would contribute $9.7 \times 
10^{-6}$ to the total energy density. This limit hence beats the BBN and CMB bounds. It also constrains
values of the cosmic strings free parameters (string tension $\mu$ and loop size $\epsilon$) not excluded
by any other experiments; for a rather low value of the string reconnection probability $10^{-3}$, $7 \times 
10^{-9} < G\mu < 1.5 \times 10^{-7}$ for $\epsilon < 8 \times 10^{-11}$. Measurements from Virgo, LIGO Hanford and LIGO Livingston probe the existence of a SGWB at higher frequency where the
Hanford-Livingston pair sensitivity is slightly worse. An upper limit on $\Omega_3 < 0.33$ at 900 Hz has 
been set combining LIGO Hanford, LIGO and Virgo pairs \cite{Abadie:2011fx}. This improves a limit set by 
LIGO Livingston and the ALLEGRO bar detector ($\mathrm{\Omega_{GW}(915 Hz) < 1.02})$~\cite{Abbott:2007wd}.

Finally, searches for point-like or more extended sources have been conducted over the same data, 
producing upper limits on the strain amplitude in each 0.25 Hz wide frequency bin.
Three specific directions (the nearby low-mass X-ray binary Sco X-1, the galactic center and SN 1987A) 
have been studied, but no indication of a SGWB has been found~\cite{Abbott:2011rr}.

\section{Towards the second generation of detectors}

In a few years from now a second generation of GW detectors 
will come online with initial improvements in sensitivity of about a factor of two with respect the the S6/VSR2-3 runs and ramping up over the space of a few years to sensitivities about of factor of ten higher \cite{Harry:2009zz}. The network will include two LIGO detectors at Hanford and Livingston, USA, a Virgo detector in Cascina, Italy, and the  smaller scale GEO detector in Germany. By the beginning of the next decade  one of the LIGO detectors is likely to become operative in India and soon thereafter a large scale detector is expected in Japan enhancing the sky coverage and source localization capabilities of the global network. 


The science runs of the second generation detectors promise to yield the first direct detection of GWs by 2020. With an order of magnitude better sensitivity, and despite neutron star merger rate uncertainties, it is 
likely that the first astrophysical events observed be one of the
1 -- 400 neutron star mergers predicted per year \cite{ratesdoc}. Search pipelines aimed at detecting such signals have been exercised in blind injection challenges \cite{Abadie:2011nz} and have demonstrated their readiness
to confidently detect such events. However there may be surprises and the broad variety of GW searches that have been developed in the past years reflects the breadth of the net that we cast. 
With the first rapid EM follow-ups of GW triggers and with the GW searches triggered by EM and neutrino observations we have taken the first steps towards multi-messenger astronomy which will be the new paradigm for the study of compact objects soon after the first few GW detections. A very exciting decade lies ahead !









\section{Acknowledgements}
The authors acknowledge their colleagues of the LIGO Scientific Collaboration and of Virgo with whom the GW results reviewed in this paper were produced. Our regular interactions through the collaborative efforts within the various working groups has certainly influenced the way that the authors have presented the material in this review.

\bibliographystyle{elsarticle-num}
\bibliography{biblio}







\end{document}